\documentclass{vgtc}                          % final (conference style)
\graphicspath{{figures/}{pictures/}{images/}{./}} % where to search for the images

\usepackage{times}                     % we use Times as the main font
\usepackage{tabu}                      % only used for the table example
\usepackage{booktabs}                  % only used for the table example
\usepackage{lipsum}                    % used to generate placeholder text
\usepackage{mwe}                       % used to generate placeholder figures

\usepackage{mathptmx}                  % use matching math font
\usepackage{siunitx}
\usepackage{gensymb} 
\usepackage{ulem}
\onlineid{0}

\vgtccategory{Research}

\vgtcinsertpkg
\usepackage{orcidlink}
\definecolor{mycolor}{RGB}{49,112,12}
\title{No More Head-Turning: Exploring Passthrough Techniques for Addressing Rear Interruptions from the Front in VR}

\author{Zixuan Guo \orcidlink{0000-0002-0451-8988}\\ %
       \parbox{1.4in}{\scriptsize \centering Xi'an Jiaotong-Liverpool University \\ The University of Liverpool} %
\and Yuekai Shi \\ %
    \parbox{1.4in}{\scriptsize \centering Xi'an Jiaotong-Liverpool University} %
\and Tiantian Ye \\ %
    \parbox{1.4in}{\scriptsize \centering Xi'an Jiaotong-Liverpool University} %
\and Tingjie Wan \orcidlink{0009-0003-0237-9587}\\ %
    \parbox{1.4in}{\scriptsize \centering Xi'an Jiaotong-Liverpool University \\ The University of Liverpool} %
\and Hai-Ning Liang~\orcidlink{0000-0003-3600-8955}\thanks{Corresponding author (e-mail: hainingliang@hkust-gz.edu.cn)}\\ %
    \parbox{1.4in}{\scriptsize \centering The Hong Kong University of Science and Technology (Guangzhou)}}

\author{
Zixuan Guo\textsuperscript{1}~\orcidlink{0000-0002-0451-8988}\thanks{This work was partially conducted while the author was affiliated with Xi'an Jiaotong-Liverpool University and the University of Liverpool.}
\and 
Yuekai Shi\textsuperscript{2}~\orcidlink{0009-0009-0122-996X}
\and 
Tiantian Ye\textsuperscript{2}~\orcidlink{0009-0002-2759-2934}
\and 
Tingjie Wan\textsuperscript{1}~\orcidlink{0009-0003-0237-9587}
\and 
Hai-Ning Liang\textsuperscript{1}~\orcidlink{0000-0003-3600-8955}\thanks{Corresponding author (e-mail: hainingliang@hkust-gz.edu.cn)}
}

\affiliation{\scriptsize \textsuperscript{1} Computational Media and Arts Thrust, The Hong Kong University of Science and Technology (Guangzhou), Guangzhou, China\\
\textsuperscript{2} School of Advanced Technology, Xi'an Jiaotong-Liverpool University, Suzhou, China}

\teaser{
  \centering
  \includegraphics[width=\linewidth]{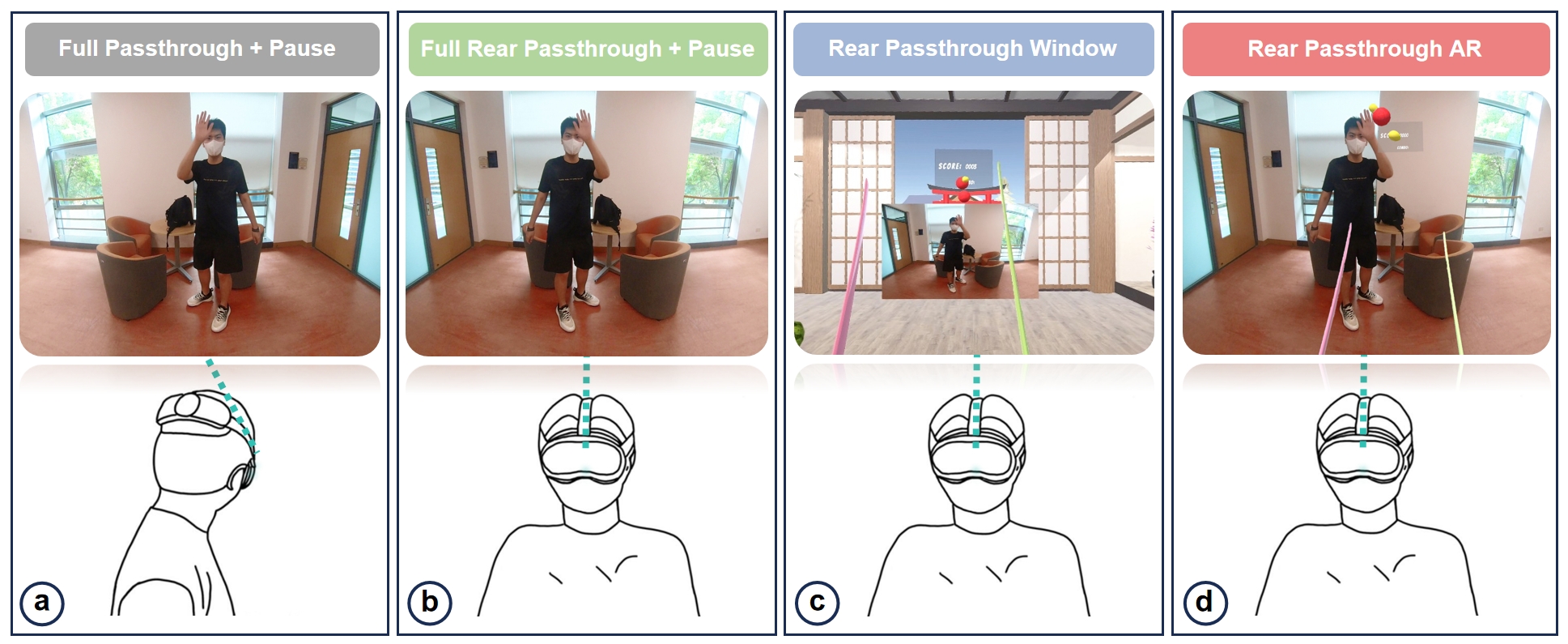}
  \caption{(a) When VR users encounter interruptions from the rear, they need to turn their heads to use Passthrough to view the real world. In this study, we propose three Passthrough techniques that display rear awareness in front of the user to help avoid head turning: (b) Full Rear Passthrough + Pause (FRPP), (c) Rear Passthrough Window (RPW), and (d) Rear Passthrough AR (RPAR).}
  \label{fig:teaser}
}

\abstract{
Virtual reality (VR) users often encounter interruptions, posing challenges to maintaining real-world awareness during immersive experiences. The Passthrough feature in VR headsets allows users to view their physical surroundings without removing the headset. However, when interruptions come from the rear, users need to turn their heads to see the real world, which can lead to negative experiences in VR. Study 1, conducted through semi-structured interviews involving 13 participants, found that users are less likely to use Passthrough for rear interruptions due to large head-turning movements, which cause inconvenience, increase the risk of motion sickness, and reduce the experience. Building on these findings, we introduced three Passthrough techniques in Study 2 for displaying the rear view in front of the user: Full Rear Passthrough + Pause (FRPP), Rear Passthrough Window (RPW), and Rear Passthrough AR (RPAR). Compared to the Baseline method that requires head-turning, all three systems reduced physical and temporal demands, alleviated disorientation caused by motion sickness, and provided a better user experience for managing rear interruptions. Among these, FRPP and RPAR were the most preferred. These findings provide valuable insights for future VR design, emphasizing the need for solutions that effectively manage rear interruptions while maintaining user comfort and experience.
} % end of abstract


\begin{document}


\firstsection{Introduction}

\maketitle

In everyday life, virtual reality (VR) users often face interruptions, be they from bystanders, objects, or pets. While VR offers complete immersion, it also isolates users from their real environment, making interruptions a critical area of research \cite{o2023re, o2020reality, o2022exploring, george2019should, wang2022realitylens}. \textcolor{black}{Previous studies have proposed various methods to help users manage real-world interruptions, including text notifications \cite{ghosh2018notifivr, o2020reality, rzayev2019notification}, audio notifications \cite{o2020reality, o2022exploring}, haptic notifications \cite{george2018intelligent, ghosh2018notifivr}, avatar designs \cite{gottsacker2021diegetic, medeiros2021promoting, kudo2021towards, george2020seamless}, symbol designs \cite{medeiros2021promoting, kudo2021towards}, and Passthrough functions \cite{gottsacker2021diegetic, von2019you, mcgill2015dose, george2020seamless, guo2024exploring}. Evaluations of these technologies reveal distinct advantages across different approaches, but, in general, participants tend to favor visual cues \cite{o2023re, von2019you, guo2024enhancement, mcgill2015dose}. Among these, Passthrough-based technologies are particularly reliable, providing better context, more accurate real-world visuals, and detailed information \cite{o2023re, guo2024enhancement, wang2022realitylens}.} In fact, Passthrough is the most commonly used method for handling interruptions in real-life situations, utilizing the headset's cameras \textcolor{black}{to offer a live video feed of} the physical environment without removing the headset \cite{kuo2023reprojection, banquiero2023passthrough, pointecker2022bridging, feld2024simple}.

% Among these, Passthrough-based technologies are considered particularly reliable, as they provide real-world visuals and convey detailed information \cite{o2023re, guo2024enhancement, wang2022realitylens}. In fact, Passthrough is the most commonly used method for handling interruptions in real-life situations. This feature uses the headset’s built-in cameras to provide a live stereo video feed of the real world, allowing users to see  a live video feed of their physical surroundings without removing the headset \cite{kuo2023reprojection, banquiero2023passthrough}. Passthrough seamlessly integrates the real world with the virtual environment, helping users stay aware of their surroundings while remaining immersed in VR \cite{pointecker2022bridging, feld2024simple}.

Interruptions can occur from all directions—the front, sides, or rear—and users' experiences with Passthrough vary accordingly. \textcolor{black}{However, previous research often overlooks this, assuming that interruptions come only from the front, which is not always realistic \cite{o2023re, guo2024enhancement, wang2022realitylens, mcgill2015dose, o2022exploring2}.} When interruptions occur in front, users can naturally look ahead without needing to turn their heads, allowing for a relatively seamless and effortless interaction between the virtual and real worlds. For interruptions on either side, users can make small head turns to check their surroundings with minimal effort, generally maintaining comfort. However, monitoring what is happening from the rear can be more challenging, as it often requires significant head rotations, potentially disrupting the immersive experience and adding some physical strain. \textcolor{black}{The direction of the interruption may be a key factor influencing user experience and shaping how interruption-handling technologies should be designed, an area that remains underexplored in the existing literature.}

\textcolor{black}{In daily life, people are generally accustomed to turning their heads, and discomfort from head movements is typically more likely to occur with faster or more repetitive actions, although this threshold can vary across individuals \cite{lawson2016neurovestibular, graybiel1968rapid}. However, in VR, this action can become less pleasant due to the} mismatch between visual and bodily feedback, leading to discomfort, fatigue, and motion sickness, which can negatively impact immersion \cite{mcgill2017passenger, cas, wang2023effect}. Furthermore, the weight of the VR headset makes head-turning more physically demanding, further contributing to discomfort \cite{walker2010head, ragan2016amplified, sargunam2017guided}. \textcolor{black}{While much of the research has focused on virtual environments, some studies \cite{santoso2024video, bailenson2024seeing, freiwald2018camera} also note that Passthrough, with its lower fidelity compared to human vision—such as limited field of view, distortion, latency, and resolution—can introduce similar risks of discomfort and motion sickness. }

Users need to adopt different approaches to handle interruptions from different directions, which may affect their use of Passthrough to view the real world, yet there is currently limited research on this aspect. To address this gap, we conducted Study 1 with 13 participants through semi-structured interviews. The results showed that participants frequently chose to use Passthrough to view the real world in response to interruptions, particularly when dealing with unidentifiable people or events, as well as situations requiring visual information. However, they were less likely to use Passthrough for rear interruptions due to the large head movements required. This underscores the need for solutions that effectively manage rear interruptions while maintaining user comfort.

Building on these insights, we proposed three Passthrough techniques to provide rear awareness in front of the user: Full Rear Passthrough + Pause (FRPP), Rear Passthrough Window (RPW), and Rear Passthrough AR (RPAR) (see Figure \ref{fig:teaser}). These techniques are designed to allow users to view their rear perpective without significant head rotation, thus minimizing physical effort and potential discomfort while also improving efficiency. The views in all three techniques are mirrored to align with users' everyday habit of using rear-view mirrors. We then conducted Study 2 to assess the effectiveness of these three techniques in helping users check rear interruptions. The study involved 24 participants and compared these three techniques with the commonly used Full Passthrough + Pause (Baseline), forming the four experimental conditions.

\textcolor{black}{The results demonstrated that all three techniques we tested improved user experience and reduced motion sickness to some extent compared to the Baseline head-turning condition, highlighting the feasibility and advantages of displaying rear awareness in front of the user. We found that participants preferred FRPP and RPAR. While previous research on front awareness \cite{o2023re, gottsacker2021diegetic, von2019you} has shown that users generally dislike pausing the application, participants appeared more open to pausing when dealing with rear awareness due to its unconventional perspective. Furthermore, based on insights from the existing literature, this work provides design recommendations for applying these techniques in real-world usage scenarios.} Overall, our paper makes the following contributions:

\begin{itemize} 
\item An investigation of how the direction of interruptions impacts VR users' utilization of Passthrough. 
\item The proposal and evaluation of three Passthrough techniques designed to present rear awareness in front of users to better manage rear interruptions. 
\item Design insights for effectively displaying rear Passthrough views from the front in VR. 
\end{itemize}

% Despite the availability of Passthrough, users do not always activate it in response to interruptions. Research has investigated the factors influencing users' decisions to use Passthrough in response to interruptions. Key factors include the importance of the interruption, the intensity of the VR activity, and the user's location. Additionally, studies suggest that the location of the interruption also plays a significant role; when interruptions occur within the gaming area, users feel a greater need to use Passthrough to check their surroundings.

\section{Related Work}

\subsection{Interruptions During VR Usage}
Due to the immersive and enclosed nature of VR headsets, interruptions have become a significant topic of research in VR. Kern et al. \cite{kern2003context} proposed five factors that define a user's interruptibility: the importance of the interruption event, the activity of the user, the social activity, the social situation, and the physical location. However, Ghosh et al. \cite{ghosh2018notifivr} observed that most VR interactions are highly personal and are not visible to others, making some of these factors less relevant. They suggested that the key factors influencing interruptions in VR environments are the importance of the event, the nature of the VR task, and the location—both physical (such as private versus shared spaces) and virtual.

Research \cite{o2023re, o2020reality, o2022exploring, george2019should, wang2022realitylens, guo2024exploring, gottsacker2022exploring, dao2021bad} has shown that interruptions in daily life often arise from interactions with bystanders. Ghosh et al. \cite{ghosh2018notifivr} found that VR experiences are frequently disrupted when bystanders enter the user's play area or attempt to interact with them. O'Hagan et al. \cite{o2021safety}, through a study of in-the-wild interactions between VR users and bystanders, found that bystanders typically interrupt users through voice or touch, seeking to engage with them. 

In addition, interruptions can come from objects and pets \cite{ghosh2018notifivr, o2023re, wang2022realitylens, hartmann2019realitycheck, kanamori2018obstacle, xiong2024petpresence, dao2021bad}. For instance, users encounter objects or hear sounds from objects that draw their attention away from the virtual world. Pets, in particular, can be an unpredictable source of interruption, as they may enter the play area, make noise, or seek the user’s attention. In summary, such interruptions prompt users to assess the urgency and relevance of real-world events against the ongoing VR task, requiring them to decide whether to gain real-world awareness to respond accordingly.

\subsection{Adapting to Interruptions with Passthrough}
To respond to real-world interruptions, users previously had to remove their headsets, but now they could utilize the Passthrough feature, which has become an integral part of VR systems. This functionality enables users to remain immersed in a virtual environment while observing their physical surroundings \cite{guo2024breaking, kudry2023enhanced}. By using the headset's built-in cameras to capture real-time stereoscopic video, Passthrough allows users to view a live video feed of their environment while using VR, helping to bridge the gap between the virtual and real worlds \cite{kuo2023reprojection, banquiero2023passthrough}.

Passthrough is typically implemented in two methods: Full Passthrough and Passthrough Augmented Reality (AR). Full Passthrough allows users to temporarily switch their entire view from the virtual world to the real world, usually activated by tapping the side of the headset or pressing a designated button \cite{gottsacker2021diegetic}. It is important to note that this method means that the ongoing VR application is paused. In contrast, Passthrough AR selectively integrates the real world into the virtual environment by retaining key virtual elements while replacing non-essential parts with a real-world view \cite{guo2024breaking, o2023re}. This approach enables users to interact with virtual content while simultaneously engaging with their physical surroundings. 

Some studies \cite{wang2022realitylens, guo2024enhancement, von2019you} have explored presenting Passthrough views through windows. For instance, Wang et al. \cite{wang2022realitylens} developed RealityLens, which lets users view the physical world through a circular window in VR. Similarly, Guo et al. \cite{guo2024enhancement} used a square window to display bystanders in shared spaces. Like Passthrough AR, these methods allow users to engage with VR while observing the real world.

Despite its effectiveness in managing interruptions, there is limited research on how Passthrough helps users respond to interruptions from different directions. Previous studies \cite{o2023re, wang2022realitylens, hartmann2019realitycheck, gottsacker2021diegetic} have focused primarily on interruptions from the front. While Passthrough works well for viewing the front or sides with minimal head movement, checking the rear requires a significant head turn, which can be more challenging in VR.

\subsection{Challenges of Head Rotation in VR} \label{challenges}
VR systems enable users to explore both virtual and real environments through natural head movements, with modern devices supporting six degrees of freedom (DOF) head tracking, including rotational and linear movements \cite{wu2019effect, keshavarz2011axis}. While this enhances immersion, it can also be physically demanding, especially with large or frequent head rotations \cite{langbehn2019turn, ragan2016amplified}. 

The added weight of VR headsets makes head rotations inconvenient and tiring. Research by Walker et al. \cite{walker2010head} found that users reduce head movement in VR, spending 59\% of their time with minimal motion. Similarly, Ragan et al. \cite{ragan2016amplified} observed that when continuous rotations are required, users often prefer virtual navigation techniques, like thumbstick control, over physical head rotations due to convenience and reduced effort.

Head rotations in VR can also cause motion sickness. Studies by Palmisano et al. \cite{palmisano2023differences} and Ujike et al. \cite{ujike2004effects} found that all types of head rotations can trigger motion sickness, particularly when rotation speeds exceed 30 to 60 degrees/second, due to discrepancies between physical movement and motion-to-photon latency \cite{warburton2023measuring, wang2022real}.

Moreover, head rotations can disrupt user experiences, as noted by Livatino et al. \cite{livatino2022effects}, causing visual disturbances that hinder task performance in dynamic settings. Cobb \cite{cobb1999measurement} also highlighted that head rotations can lead to postural instability, affecting activities such as simulated driving or machinery operation.

In general, head rotations in VR present notable challenges. While most previous studies \cite{barai2020outside, nwobodo2023review, ragan2016amplified, palmisano2023differences} have focused on head rotations within the virtual environment, \textcolor{black}{some studies \cite{santoso2024video, bailenson2024seeing, freiwald2018camera, wang2024omnidirectional} have highlighted that Passthrough, due to its lower fidelity compared to natural vision—characterized by factors like limited field of view, distortion, latency, and resolution—can also pose risks of discomfort and motion sickness. These factors can also make turning the head while using Passthrough challenging. Yet few studies have explored how interruptions from different directions in the real world influence users to utilize Passthrough as a response.}

\section{Study 1: Exploring VR Users' Passthrough Use for Addressing Interruptions and Directional Impacts}
\textcolor{black}{Many studies (e.g., \cite{o2023re, mcgill2015dose, o2022exploring2}) have highlighted the benefits of using Passthrough to handle interruptions. However, few have focused on the impact of interruption direction, leaving a gap in our understanding of how Passthrough is used in real-life scenarios. To address this gap, we conducted Study 1, using semi-structured interviews to examine the influence of interruption direction.} We first explored participants' experiences with Passthrough in managing interruptions and then investigated how interruptions from different directions influenced their use of Passthrough. 

\subsection{Participants, Apparatus, and Procedure}

We recruited 13 participants (\textcolor{black}{8 females, 5 males}; mean age = 25.3, SD = 4.27, range = 21 to 33) through social media platforms, ensuring data saturation \cite{guest2006many}. All participants had prior experience with VR; 9 used VR 3-4 times per week, while 4 used it at least once per week. To help participants simulate the experience of interruptions during VR usage, \textcolor{black}{we provided a wireless PICO 4 VR headset.}

% Participants could switch to a full Passthrough view by tapping the side of the headset.

The study took place in a laboratory on the university campus. After signing the consent form, participants were asked about their demographics and VR usage experience. They were then asked about their experiences using Passthrough to handle real-world interruptions while using VR. \textcolor{black}{Specifically, they were asked: \textit{What types of interruptions have you experienced while using VR? Do you use Passthrough to handle these interruptions? If so, how do you use it? If not, why?}}

Following this, participants were provided with a VR headset, and the experimenter simulated interactions around them based on the examples provided in the previous phase. \textcolor{black}{During this process, participants were immersed in a default Unity skybox environment. For each example (e.g., attempting to initiate a conversation with the participant), the experimenter simulated an interaction in front of, to the side of, and to the rear of the participant. The interactions began 20 seconds after the participant had been immersed in the virtual environment, with a 20-second interval between each interaction. Participants were asked: \textit{Does the direction of the interruption influence your willingness to use Passthrough to check it?} If they answered yes, they were then asked: \textit{What is your preference, and why?} If they answered no, they were further asked to explain the reason. During the experiment, we did not impose any restrictions on the participants' head-turning movements, as this is a natural posture that depends on their personal habits. Some participants mainly rotated their heads, while others slightly turned their bodies before rotating their heads, but both methods resulted in the same angle of head rotation to view the rear.} Each interview lasted approximately 20 minutes and was fully recorded on video. 

\textcolor{black}{We adopted a thematic analysis combined with a quantitative approach to report the frequency of mentions, providing additional insights into the themes raised by participants. According to a review by Bowman et al. \cite{bowman2023using} on the use of thematic analysis methods in HCI, this mixed approach is commonly used in the field of HCI (e.g., \cite{coutrix2024impact, kim2022mobile}). }Two coders independently coded the same five interviews. They then discussed the identified themes and agreed on a coding scheme. One coder used this scheme to code all the interviews. \textcolor{black}{To ensure reliability, after completing the coding, the second coder randomly selected two interviews from the set to review the coding. Any discrepancies between the coders were discussed and resolved.}

\subsection{Results}
\textcolor{black}{The semi-structured interview results revealed participants' experiences with using Passthrough to address interruptions, including their reasons for using it and how they interacted with it. Although previous research \cite{gottsacker2021diegetic, o2023re, medeiros2021promoting} has explored Passthrough usage for interruptions, few have examined how the direction of interruptions influences its use. Our findings fill this gap, highlighting how the direction of interruption affects users' willingness to engage with Passthrough.}

\subsubsection{Experiences of Using Passthrough to Address Real-World Interruptions}
Participants primarily reported interruptions during VR use related to \textbf{interactions with bystanders}. 9 participants mentioned bystanders attempting to initiate conversations, 5 noted someone suddenly entering the room, and 4 recalled bystanders talking, laughing, or making noise nearby. Furthermore, some participants described \textbf{interactions with objects} and \textbf{interactions with pets}—3 mentioned encountering obstacles, 2 mentioned that certain sounds, like something falling, caught their attention, and 2 noted pets running or barking nearby.

The process of handling interruptions using Passthrough was generally similar for all 11 participants. Upon noticing an interruption, they would first \textbf{switch to Passthrough to observe the real world}. Participants then \textbf{evaluated the interruption} to assess its seriousness. Based on their evaluation, they would \textbf{take appropriate actions}: If the event was too serious or could not be resolved quickly, they would remove the VR headset to address the situation fully. For less urgent interruptions, they would respond directly within the Passthrough view, maintaining their immersive experience. 

We found that participants mainly used Passthrough to view interruptions involving unidentifiable people or events, as well as situations requiring visual information. 5 participants mentioned the need to \textbf{identify bystanders}. For example, P1 and P8 mentioned, \textit{“When I’m playing alone and hear someone suddenly enter the room, I use Passthrough to see who it is.”} Furthermore, 5 participants provided examples \textbf{involving unfamiliar bystanders}. As P12 explained, \textit{“When I hear a stranger calling me, I stop to see what he wants to say.”} When faced with similar interruptions from familiar individuals, they generally felt more at ease and did not necessarily need to visually perceive the situation. Furthermore, 6 participants expressed a need to \textbf{investigate unidentifiable events}. For instance, P6 shared, \textit{“I heard someone laughing and thought they were talking about me, so I turned on Passthrough and saw they were watching a funny video together.”} Similarly, P4 mentioned, \textit{“When I hear my pet making noise, I need to see what’s happened.”} P7 added, \textit{“I suddenly hear something falling and making a loud noise, I stop to check what's going on.”} Additionally, there are situations where users are aware of what is happening, but the interruptions still \textbf{require visual confirmation} to be resolved. As P12 mentioned, \textit{“When I suddenly bump into furniture, I need to turn on Passthrough to check my surroundings.”} P5 noted, \textit{“When visual confirmation is needed during a conversation, like when my mom asks how her outfit looks, I have to switch to Passthrough.”}

\subsubsection{Impact of Interruption Direction on the Use of Passthrough}
8 participants indicated that the direction of an interruption affects their use of Passthrough to view the real world. They reported the highest preference for using Passthrough for interruptions from the front, followed by the sides, and the least for interruptions from the rear. When asked for reasons, 7 participants mentioned that turning their heads more than 90° felt \textbf{bothersome and inconvenient}. P11 mentioned, \textit{“A slight turn to check something on the side is acceptable, but turning completely around is strenuous.”} P3 shared, \textit{“I might be curious about what’s happening, but if I have to turn around to look and then turn back to continue playing, it feels like too much trouble, so I’d rather not look.”} 4 participants stated that making large head turns while wearing a VR headset felt \textbf{uncomfortable}, with P2 even mentioning that it could lead to dizziness or even nausea. Moreover, 3 participants noted that turning to check the rear \textbf{disrupted immersion} in the virtual environment. As P1 explained, \textit{“When I turn my head to look behind me and then turn back, it feels like I’m re-entering the virtual environment.”} 

In contrast, 5 participants stated that the direction of the interruption did not affect their use of Passthrough, mainly due to the \textbf{need to check}. However, 2 participants still expressed negative attitudes toward turning their heads. For example, P9 said,  \textit{“If I need to address an interruption, I have to check it no matter which direction it comes from. Although it is indeed more troublesome when it's behind me."} 
% Additionally, 11 participants pointed out that when checking the rear, Passthrough AR offered little value because turning their heads required pausing the VR application.

\color{black}
\subsection{Discussion}
The types of interruptions encountered by participants align with previous research \cite{o2023re, o2020reality, o2022exploring, george2019should, wang2022realitylens}, and, similarly, our findings support prior studies \cite{o2023re, o2022exploring2} that Passthrough is more suitable for less urgent situations. While some research \cite{mcgill2015dose, o2023re, guo2024breaking} has explored how Passthrough aids users in managing interruptions, our findings further emphasize the scenarios where users tend to rely on Passthrough to obtain comprehensive visual cues: specifically, when addressing unidentifiable bystanders and events or when visual confirmation is necessary.

Although many studies \cite{gottsacker2021diegetic, o2023re, o2022exploring2} have examined the impact of interruptions on VR users and strategies to help them manage such events, most have overlooked the influence of the direction of the interruption. Our research fills this gap, showing that the direction of an interruption significantly affects users' willingness to use Passthrough. Interruptions from the rear were the least preferred and considered the most challenging compared to those from the front or sides. The key reasons cited by participants were (1) the effort and inconvenience required, (2) physical discomfort, and (3) disruption to their immersion in the virtual environment. 
\color{black}

\section{Study 2: Evaluating Passthrough Techniques for Providing Rear Awareness}
\textcolor{black}{Consistent with prior research \cite{o2023re, guo2024enhancement, wang2022realitylens}, Study 1 demonstrates that users have a preference for visual cues when responding to various interruption events, with Passthrough being seen as particularly practical due to its ability to display the real world. However, we also observed that users are less inclined to use Passthrough to check behind them, as it requires the extra effort of turning their heads. } To address this, we built on previous research \cite{guo2024breaking, guo2024enhancement, o2023re, von2019you, wang2022realitylens} and proposed three systems that provide rear awareness by displaying Passthrough views in front of the user (Figure \ref{fig:teaser}). 

Considering the common use of rear-view mirrors in daily life, the Passthrough views in these three systems were designed to be mirrored. Study 2 used a within-subjects design to compare four experimental conditions, including three proposed systems and the commonly used Full Passthrough + Pause (Baseline). The order of the conditions was counterbalanced in the experiment. The details of each condition are as follows:

\begin{itemize}
\item \textbf{Full Passthrough + Pause (Baseline):} In this baseline condition, when users perceive an interruption from the rear, they activate the Full Passthrough view, pausing the VR application. They then need to turn their heads to check the event from the rear before turning back to resume their VR activity.

\item \textbf{Full Rear Passthrough + Pause (FRPP):} When interrupted from the rear, users activate the Full Rear Passthrough view, which pauses the game and displays the rear view directly in front of them. After checking, they close the view and resume their VR activity.

\item \textbf{Rear Passthrough Window (RPW):} This system presents the rear Passthrough view in a small window in front of the user, allowing them to observe the event from the rear without pausing the application. This window, measuring \SI{120}{cm} in length and \SI{45}{cm} in width, clearly displays the rear view.

\item \textbf{Rear Passthrough AR (RPAR):} In this system, key virtual elements of the application are retained while the rest of the view is replaced by the Rear Passthrough view, enabling users to observe the event from the rear without pausing the application.

\end{itemize}

\textcolor{black}{It is worth noting that in previous studies \cite{o2023re, gottsacker2021diegetic}, the full Passthrough view was always accompanied by pausing the application, as this was necessary for switching the view and was an integral part of the feature. In contrast, other methods do not require pausing the application and can provide real-world cues while the application continues. For these methods, whether or not to pause the game depends on the user's personal needs.}

\subsection{Pilot Study}
We conducted a pilot study with 8 participants (\textcolor{black}{4 females, 4 males}; mean age = 26.75, SD = 5.12, range = 21 to 32) to evaluate the design and feasibility of three Passthrough techniques that display rear views in front of users. Participants experienced both mirrored and non-mirrored versions of these systems. In this pilot study, we used a default Unity skybox as the virtual environment and a 360° photo of a room as the physical environment. The program was packaged into a PICO 4 VR headset. Overall, participants felt that all three systems effectively allowed them to observe the rear without having to turn their heads, which was viewed as a positive enhancement to their VR experience. As anticipated, 7 participants found that the mirrored view was particularly effective in determining the location of rear events, consistent with their usual use of mirrors or rear-view mirrors for better spatial awareness. Additionally, based on their feedback, we adjusted the size and position of the window in RPW to ensure that it would be easy to observe.

\subsection{VR Game}
\begin{figure}[htbp]
    \centering
    \includegraphics[width=\columnwidth]{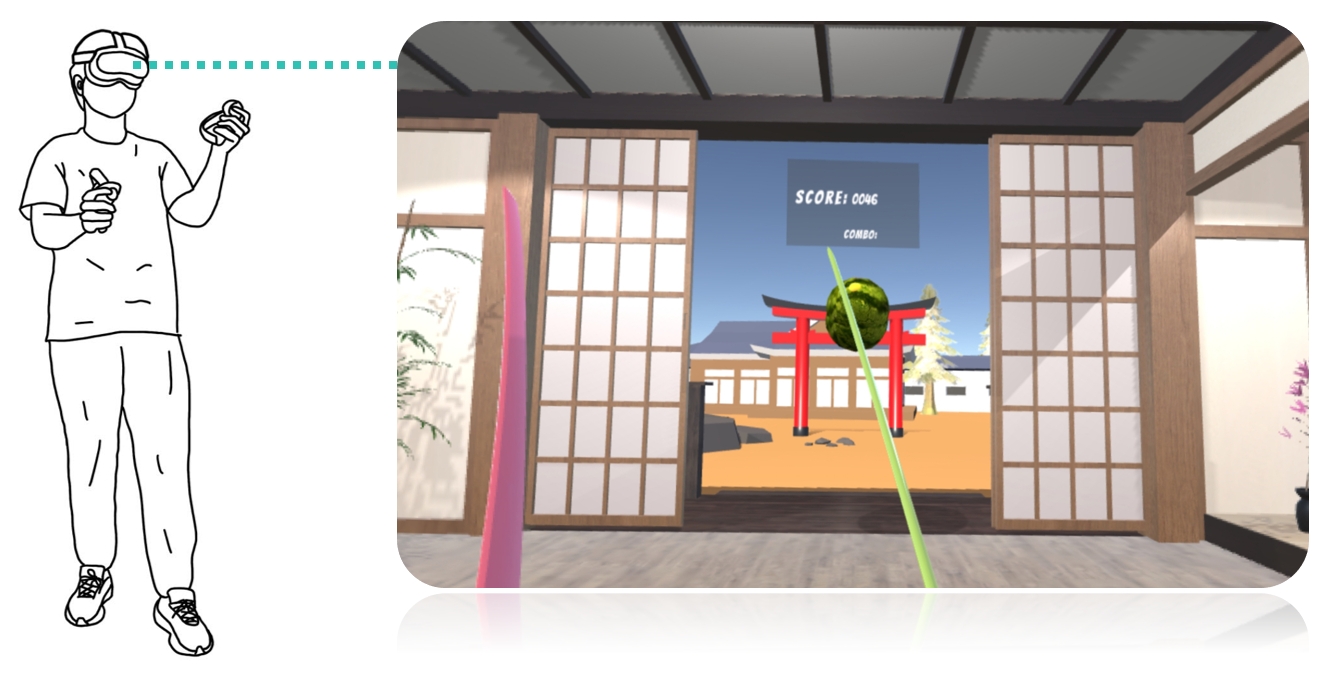}
    \caption{A participant using the controllers to slice the incoming fruits in the game.}
    \label{game}
\end{figure}

For our experiment, we used a game as the VR application scenario because it effectively represents typical VR consumer applications, consistent with the use of gaming scenarios in previous research on awareness systems \cite{o2023re, von2019you}. We created a game inspired by the VR Fruit Ninja \footnote{https://store.steampowered.com/app/486780/Fruit\_Ninja\_VR/}, set in a Japanese-style room (Figure \ref{game}). The game effectively captures attention and quickly immerses participants in the virtual environment, making it ideal for assessing user experiences that involve transitioning between virtual and real worlds \cite{guo2024enhancement}. In the game, players use two handheld controllers to control virtual swords, with the aim of cutting as many fruits (watermelon, apple, and lemon) as possible while avoiding bombs. The game mechanics are similar to those in \cite{guo2024breaking}, with fruits and bombs being launched from both the left and right sides of the player in a parabolic arc. The game is played over a period of 3 minutes. \textcolor{black}{To maintain a consistent difficulty level, we designed three one-minute sequences for generating fruits and bombs. Each sequence consists of 30 rounds, each lasting 2 seconds, during which a fixed number of fruits and bombs are generated in random order. Each round contains 2-4 fruits, with a bomb appearing every 5 rounds. There is no time gap between the sequences.}

\subsection{Interruption Events}

\begin{figure*}[htbp]
    \centering
    \includegraphics[width=\textwidth]{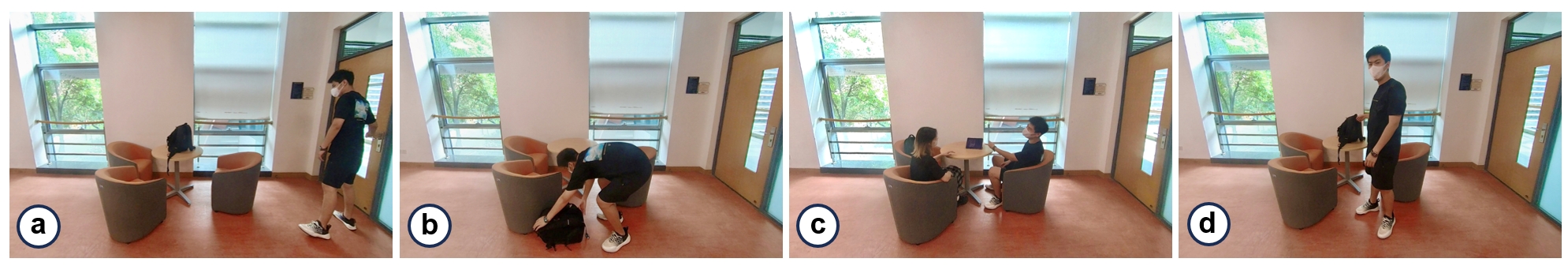}
    \caption{The four types of interruptions in the experiment: (a) Someone Entering the Room, (b) an Object Falling, (c) Someone Laughing, and (d) Someone Asking about an Object.}
    \label{events}
\end{figure*}

Based on previous research \cite{o2023re, gottsacker2021diegetic, von2019you, scavarelli2017vr, kudo2021towards}, we employed a Wizard of Oz \cite{hanington2019universal} approach, using pre-recorded video materials for the interruption events. This method ensured greater reliability and consistency in the experiences participants encountered \cite{o2023re, kudo2021towards}. All materials were filmed in a classic-layout student break room at the university, which featured some tables and chairs. We set up four types of interruption events (Figure \ref{events}) based on instances mentioned by participants in Study 1, including unidentifiable people or events, and situations requiring visual information. The experimenters appearing in the video were unfamiliar to all participants. The details of the four events are as follows:

\begin{itemize}
\item \textbf{Someone Entering the Room:} A person opens the door, walks into the room, sits down, briefly looks at the VR user, and then starts browsing the smartphone.
\item \textbf{Object Falling:} A heavy backpack falls from the table, and a person then picks it up and places it back.
\item \textbf{Someone Laughing:} Two people sit at the table, watching a funny video on a pad together and laughing loudly.
\item \textbf{Someone Asking about an Object:} A person approaches, calls out ``hey" to the user, then points to the backpack on the table and asks, ``Is this your bag?"
\end{itemize}

\textcolor{black}{Based on previous research \cite{o2023re, von2019you, mcgill2015dose, gottsacker2021diegetic}, each interruption event was triggered after the player had been playing the game for 1 or 2 minutes, ensuring that they were immersed in the game. When an interruption occurred, participants would hear the corresponding sounds: }the door opening, the backpack falling, the loud laughter, and the “hey” sound. \textcolor{black}{They were instructed to activate the Passthrough system by pressing the “A” button on the controller to check what was happening from the rear and to understand the situation.} Each condition randomly featured one of these four events.

\subsection{Measures}
\begin{itemize}

\item \textit{\textbf{Game Performance.}} We collected participants' total scores and maximum combo counts in the game as measures of their performance.

\item \textit{\textbf{System Usability.}} \textcolor{black}{Due to the unique nature of awareness systems, which balance virtual and real-world information, many studies \cite{guo2024enhancement, o2023re, o2022exploring, gottsacker2021diegetic, von2019you} have not used standard questionnaires. These traditional usability measures focus more on the general interface or interaction metrics, leading researchers to design custom questionnaires based on empirical insights and the literature to better assess their unique challenges. }Thus, we followed \cite{o2023re} to assess the system usability by asking participants to what extent they agreed with these 8 statements: (1) “was disruptive”, (2) “was frustrating”, (3) “was urgent”, (4) “felt natural”, (5) “was easy to understand”, (6) “was informative”, (7) “improved your ability to communicate with a bystander”, and (8) “made you too aware of the real world”. These statements were rated on a 7-point Likert scale.

\item \textit{\textbf{Workload.}} The workload of the awareness systems used in the experiment was measured via the NASA-TLX workload questionnaire \cite{hart2006nasa}. This tool includes 6 subscales that measure mental, physical, and temporal demand, as well as frustration, effort, and performance. Each subscale was rated on a scale from 0 to 100, with increments of 5.

\item \textit{\textbf{Presence Experience.}} We measured participants' sense of presence in the virtual environment using the ``Sense of Being There'' subscale and ``Involvement'' subscale
of the Igroup Presence Questionnaire (IPQ) \cite{schubert2001experience}, consisting of 5 items. The scale was measured on a 7-point Likert scale.

\item \textit{\textbf{Motion Sickness.}} Motion Sickness was evaluated using the Simulator Sickness Questionnaire (SSQ) \cite{kennedy1993simulator}, which consists of 16 items rated from 0 to 3. The questionnaire measures three aspects: nausea, oculomotor discomfort, and disorientation. The individual sub-scores contribute to a total SSQ score, where a range of 20 to 30 indicates mild to moderate simulator sickness, and scores above 40 suggest an unsatisfactory simulator experience \cite{caserman2021cybersickness}.

\item \textit{\textbf{Interview.}} Participants were asked to rank their preferred system for checking interruptions from the rear while using VR in the future and to explain their choices. The interviews were audio-recorded and transcribed for data analysis.

\end{itemize}

\color{black}

\subsection{Hypotheses}
Based on the challenges of head-turning in VR mentioned in previous literature (See Section \ref{challenges}), we propose the following hypotheses:

\begin{itemize}

% \item {\textbf{H1.}} The three Passthrough techniques that display rear awareness in front of the user can assist users in identifying interruptions from the rear {\textbf{(a)}} without affecting game performance or {\textbf{(b)}} reducing system usability.

% \item {\textbf{H2.}} The three Passthrough techniques that display rear awareness in front of the user can address issues identified in the Baseline, including: {\textbf{(a)}} reducing the time and physical effort required by the user, {\textbf{(b)}} minimizing disruption to immersion, and {\textbf{(c)}} alleviating motion sickness.

\item {\textbf{H1a}} The three proposed Passthrough techniques can help users identify rear interruptions without affecting game performance.

\item {\textbf{H1b}} The three proposed Passthrough techniques can help users identify rear interruptions without reducing system usability.

\item {\textbf{H2a}} The three proposed Passthrough techniques can reduce users’ time and physical effort for addressing interruptions.

\item {\textbf{H2b}} The three proposed Passthrough techniques can minimize disruption to immersion.

\item {\textbf{H2b}} The three proposed Passthrough techniques can alleviate motion sickness.

\end{itemize}

\color{black}

\subsection{Participants, Apparatus, and Setup}
We recruited 24 participants (10 females, 14 males; mean age = 28.5, SD = 8.43, range = 21 to 51) through social media platforms. \textcolor{black}{Of these, 13 had prior experience with VR headsets, with 5 using them on a weekly basis, 4 using them monthly, and 4 using them occasionally once a year.} All participants volunteered for the study without receiving compensation.

The experimental application was developed using Unity3D, version 2022.3.17f1c1. We used the PICO 4 VR headset, with the experimental application installed. Interruption events were recorded using the Insta360 ONE X2, capturing all events as 360° videos. The experiment took place in the same room where the videos had been previously recorded. The space was well-lit and isolated from any other external disturbances, and an experimenter closely supervised the session to minimize any potential risks. Ethical approval for the study was granted by the University Ethics Committee of the host institution.

\subsection{Procedure}
Upon arriving at the experiment site, i.e., the break room, participants were given an overview of the experiment's objectives and procedures. They were then provided with consent forms to review and sign. Afterward, they completed a pre-experiment questionnaire, which collected demographic information and assessed motion sickness \cite{kennedy1993simulator}.

Before the experiment began, the experimenter introduced the four awareness systems used in the study and showed a video demonstration of each. Participants were then provided with gameplay instructions and engaged in a one-minute uninterrupted training session to become familiar with the equipment and gameplay.

During the experiment, before each condition began, the experimenter assisted participants in donning the VR headset and guided them to stand in a designated spot facing a specific direction. After each condition, participants were asked to describe the interruption event that occurred from the rear. They were then asked to fill out questionnaires to evaluate the system's usability \cite{o2023re}, their perceived workload \cite{hart2006nasa}, presence experience \cite{schubert2001experience}, and motion sickness \cite{kennedy1993simulator}. At the end of the experiment, they participated in a semi-structured interview. The entire experiment took approximately 30 minutes per participant.

\begin{figure*}[htbp]
    \centering
    \includegraphics[width=\textwidth]{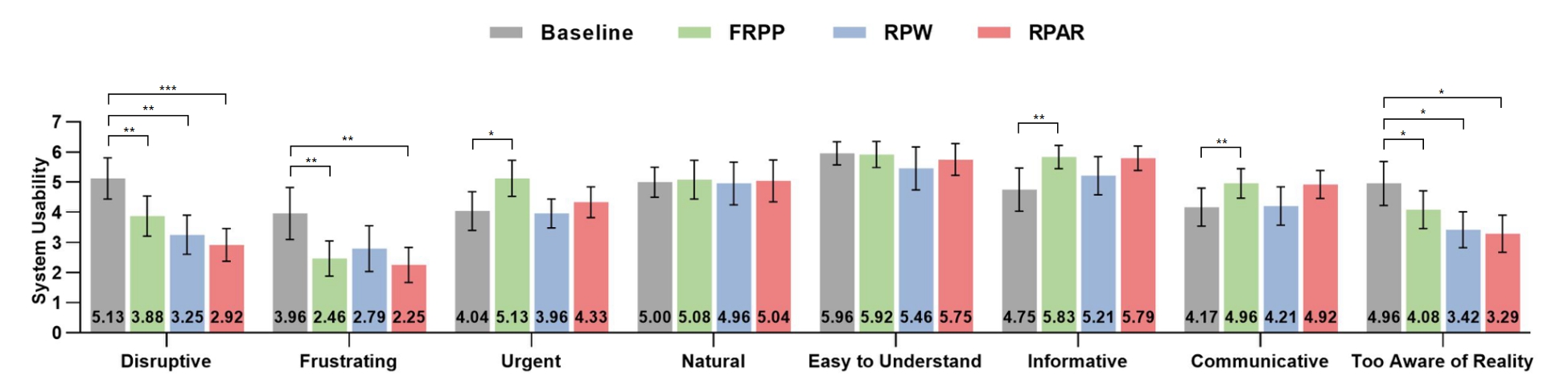}
    \caption{System usability scores for each condition, showing significant differences in Disruptive, Frustrating, Urgent, Informative, Communicative, and Too Aware of Reality. Error bars indicate 95\% confidence intervals. *, **, and *** indicate statistical significance at the $p < 0.05, p < 0.01,$ and $p < 0.001$ levels, respectively.}
    \label{Usability}
\end{figure*}

\subsection{Results}
We performed one-way repeated-measures ANOVAs (RM-ANOVA) and used Bonferroni corrections for all pairwise comparisons, reporting effect sizes ($\eta_p^2$) where applicable. Before performing the ANOVA, we checked the normality of the data using Shapiro-Wilks tests and Q-Q plots. Non-normally distributed data were transformed using the Aligned Rank Transform (ART) \cite{wobbrock2011aligned} prior to the analysis. When Mauchly’s test revealed a violation of the sphericity assumption, we adjusted the degrees of freedom using Greenhouse-Geisser estimates. \textcolor{black}{For the interview results, we used an analysis approach similar to that of Study 1, employing thematic analysis along with a quantitative statistical approach.}

\subsubsection{Performance}
Performance measures did not show significant differences across conditions. The total game scores for participants were: Baseline ($M=415.50, SD=132.75$), FRPP ($M=430.75, SD=138.54$), RPW ($M=438.87, SD=172.64$), and RPAR ($M=436.92, SD=153.16$).

\subsubsection{System Usability}

The usability evaluation results are presented in Figure \ref{Usability}. Significant differences in disruptive scores were observed between conditions ($F_{1.847,42.474}=11.825, p<.001, \eta_p^2=0.340$). The Baseline condition ($M=5.13, SD=0.33$) caused more disruption than FRPP ($M=3.88, SD=0.32$) ($p=.005$), RPW ($M=3.25, SD=0.31$) ($p=.009$), and RPAR ($M=2.92, SD=0.26$) ($p<.001$). 

Frustration levels differed significantly across conditions ($F_{3,69}=7.804, p<.001, \eta_p^2=0.253$). Post-hoc tests revealed that the Baseline condition ($M=3.96, SD=0.42$) led to significantly higher frustration compared to FRPP ($M=2.46, SD=0.28$, $p=.001$) and RPAR ($M=2.25, SD=0.28$, $p=.006$).

Further differences were found in urgency, informativeness, and communication effectiveness. For urgency ($F_{1.996,45.899}=3.859, p=.028, \eta_p^2=0.144$), FRPP ($M=5.13, SD=0.29$) generated a stronger sense of urgency than Baseline ($M=4.04, SD=0.31$) ($p=.012$). Regarding informativeness ($F=1.963,45.142=5.180, p=.010, \eta_p^2=0.184$), Baseline ($M=4.75, SD=0.35$) provided less information than FRPP ($M=5.83, SD=0.19$) ($p=.003$) and RPAR ($M=5.79, SD=0.20$) ($p=.016$). Communication effectiveness ($F_{2.178,50.103}=4.583, p=.013, \eta_p^2=0.166$) was also lower in Baseline ($M=4.17$, $SD=0.31$) compared to FRPP ($M=4.96, SD=0.24$) ($p=.001$).

Significant differences were also found in participants' tendency to be overly aware of the real world ($F_{2.015,46.338}=7.469, p=.002, \eta_p^2=0.245$), with the Baseline condition ($M=4.96$, $SD=0.35$) causing participants to be more focused on the real world compared to FRPP ($M=4.08, SD=0.30$, $p=.043$), RPW ($M=3.42, SD=0.29$, $p=.016$), and RPAR ($M=3.29, SD=0.30$, $p=.018$).

\subsubsection{Workload}
Figure \ref{Measure}.a presents the NASA-TLX scores for the four conditions. We observed significant differences in overall task workload ratings across the conditions ($F_{1.670,38.415}=18.958, p<.001, \eta_p^2=0.452$). Post-hoc pairwise comparisons indicated that participants experienced a higher workload in the Baseline condition ($M=24.83, SD=2.75$) compared to FRPP ($M=14.97, SD=1.88$) ($p<.001$), RPW ($M=19.76, SD=3.12$) ($p=.001$), and RPAR ($M=18.23, SD=2.97$) ($p<.001$).

For NASA-TLX workload subscales, we found significant differences in perceived mental demand across the four conditions ($F_{2.187,50.298}=7.894, p=.001, \eta_p^2=0.256$). Post-hoc tests revealed that participants experienced lower mental demand in the Baseline condition ($M=13.96$, $SD=2.57$) compared to RPW ($M=25.42, SD=4.51$) ($p=.012$) and RPAR ($M=24.58, SD=4.69$) ($p=.021$). 

We also found significant differences in participants' perceived physical demands across the four conditions ($F_{3,69}=15.717, p<.001, \eta_p^2=0.406$). Participants reported needing a higher physical effort in the Baseline condition ($M=30.00, SD=4.93$) compared to FRPP ($M=9.58, SD=2.69$) ($p<.001$), RPW ($M=17.92, SD=4.53$) ($p=.002$), and RPAR ($M=16.04, SD=4.45$) ($p=.001$). 

Significant differences in time demands were found ($F_{1.640,37.726}=15.461, p<.001, \eta_p^2=0.402$), with participants feeling higher time pressure in Baseline ($M=27.71, SD=3.20$) compared to FRPP ($M=18.13, SD=2.08$) ($p=.001$), RPW ($M=15.83, SD=2.27$) ($p=.003$), and RPAR ($M=13.75, SD=2.07$) ($p<.001$). 

Additionally, significant differences were also observed in perceived effort ($F_{1.868,42.968}=6.485, p=.004, \eta_p^2=0.220$), with the Baseline condition ($M=28.96$, $SD=3.83$) requiring more effort than FRPP ($M=20.63, SD=3.34$) ($p=.001$).

\subsubsection{Presence Experience}
\begin{figure*}[htbp]
    \centering
    \includegraphics[width=\textwidth]{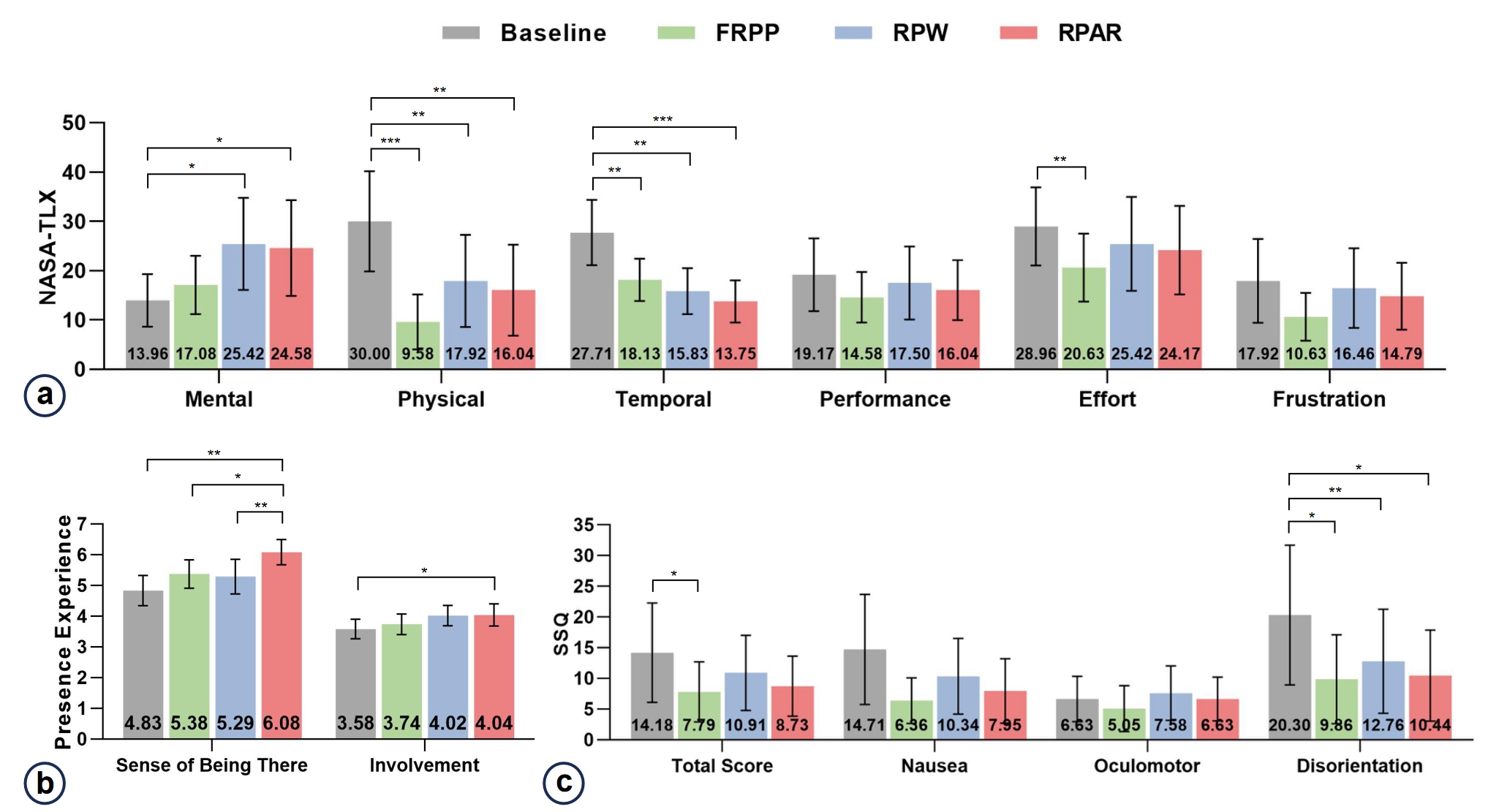}
    \caption{(a) NASA-TLX scores for each condition, showing significant differences in Mental Demand, Physical Demand, Temporal Demand, and Effort. (b) Presence experience scores for each condition, showing significant differences in Sense of Being There and Involvement. (c) SSQ scores for each condition, showing significant differences in Total Score and Disorientation. Error bars indicate 95\% confidence intervals. *, **, and *** indicate statistical significance at the $p < 0.05, p < 0.01,$ and $p < 0.001$ levels, respectively.}
    \label{Measure}
\end{figure*}

Participants' presence experiences are shown in Figure \ref{Measure}.b. A significant difference in the sense of being there was observed between conditions ($F_{2.109,48.512}=7.779, p=.001, \eta_p^2=0.253$). Post-hoc analysis indicated that participants felt a stronger presence in RPAR ($M=6.08, SD=0.20$) compared to Baseline ($M=4.83, SD=0.24$) ($p=.001$), FRPP ($M=5.38, SD=0.22$) ($p=.031$), and RPW ($M=5.29, SD=0.27$) ($p=.004$). 

For involvement, significant differences were found across conditions ($F_{3,69}=5.667, p=.002, \eta_p^2=0.198$). Participants felt more involved in RPAR ($M=4.04, SD=0.18$) than in Baseline ($M=3.58, SD=0.15$) ($p=.028$).

\subsubsection{Motion Sickness}
Figure \ref{Measure}.c presents the SSQ scores for the four conditions. There were significant differences in total SSQ scores between the conditions ($F_{1.892,43.516}=6.343, p=.004, \eta_p^2=0.216$). Post-hoc comparisons demonstrated that motion sickness was notably higher in the Baseline condition ($M=14.181, SD=3.92$) compared to FRPP ($M=7.79, SD=2.36$)  ($p=.013$). 

Furthermore, we observed significant differences in disorientation score across the conditions ($F_{1.742,40.063}=7.923, p=.002, \eta_p^2=0.256$). Participants reported significantly more disorientation in the Baseline condition ($M=20.30, SD=5.49$) than in FRPP ($M=9.86, SD=3.50$)  ($p=.010$) and RPAR ($M=10.44, SD=3.58$)  ($p=.031$).

\subsubsection{Interview Results}

\begin{figure}[htbp]
    \centering
    \includegraphics[width=0.93\columnwidth]{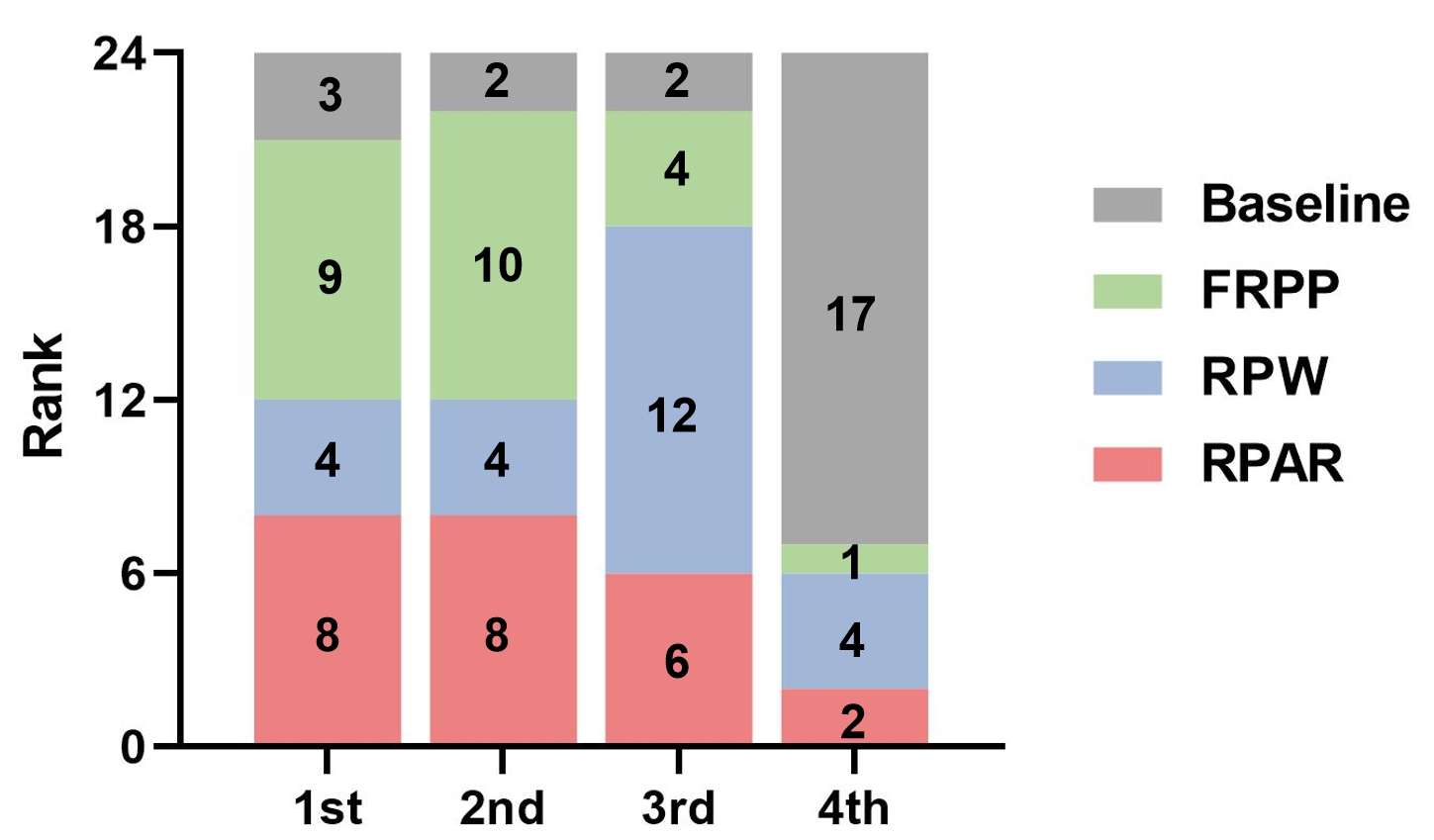}
    \caption{Participant preference ranking across the four conditions.}
    \label{rank}
\end{figure}

As shown in Figure \ref{rank}, FRPP emerged as the most preferred option among participants, followed by RPAR, then RPW, and lastly, Baseline. A total of 9 participants ranked FRPP as their top choice, followed by 8 for RPAR, 4 for RPW, and 3 for Baseline. When considering their second choice, 10 participants again favored FRPP, 8 selected RPAR, 4 chose RPW, and 2 opted for Baseline.

Regarding FRPP, 8 participants appreciated the pause feature, finding it \textbf{easy to observe}, while 3 participants valued how it \textbf{minimized distractions}. For instance, P8 noted, \textit{“Pausing helps prevent distractions, reduces mental stress, and allows better focus on real-world events.”} However, 5 participants felt the pause \textbf{disrupted}. As P2 explained, \textit{“While it's quick to check behind, it takes time to readjust when returning to the game.”}

For RPAR, 7 participants described it as \textbf{non-intrusive}, 4 appreciated its \textbf{clear presentation}, 4 found it \textbf{convenient}, and 3 thought it was \textbf{interesting}. For instance, P1 remarked, \textit{“It doesn't disrupt the game, and it fully displays the real world, making it easier to see what's happening.”} However, 4 participants found the large real-world display \textbf{distracting}. As P15 noted, \textit{“There’s too much information, and it’s hard to manage everything at once.”} However, 3 participants mentioned that they might become accustomed to this system with long-term use.

As for RPW, 4 participants liked that it did not pause gameplay and \textbf{non-intrusive}. P19 noted, \textit{“It impacts the game less than RPAR, allowing me to glance at the real world while still playing.”} However, 9 participants found that the window \textbf{required more attention}. P11 shared, \textit{“The window is small, and I need to focus on it, which disrupts my gameplay.”} Additionally, 4 participants felt the window's design did not blend well with the game, \textbf{reducing immersion}. P6 mentioned, \textit{“It feels like a floating video window on a phone, which feels disconnected from the game and diminishes immersion.”}

For the Baseline condition, 3 participants felt it \textbf{matched their usual habits}, and another 2 found it to be more \textbf{adaptable}. For example, P13 stated, \textit{“I'm not used to those three methods, and I need some time to figure out whether what I'm seeing is coming from the front or the back,”} while P16 added, \textit{“Turning around to face someone when speaking to them shows respect.”} Overall, most participants appreciated the three proposed Passthrough techniques that enable rear-viewing without the need for head-turning. Their dissatisfaction with head-turning to check rear interruptions aligns with the findings from Study 1.
% 11 participants found head-turning \textbf{bothersome and inconvenient}, 6 participants reported \textbf{experiencing dizziness}, 4 found it \textbf{time-consuming}, and 4 felt that the interruption and head turn significantly \textbf{disrupted immersion}.

\subsection{Discussion}
\textcolor{black}{Our results support \textbf{H1a}, as we found that the three Passthrough techniques, which display rear views in front of the user, had an impact on participants' game performance compared to the Baseline. Furthermore, our results also support \textbf{H1b}, showing that these three techniques not only did not reduce usability compared to the Baseline, but actually improved usability by minimizing distractions and helping users stay focused on the virtual environment without becoming overly aware of the real world. }Head-turning in VR can be disruptive and prolong exposure to the physical environment, potentially drawing excessive attention away from the virtual experience. Similarly, some research \cite{guo2024enhancement, o2023re} suggests that while Passthrough provides real-world awareness, it can lead users to pay more attention to the physical environment, thereby diminishing the sense of immersion in the virtual environment. 

\textcolor{black}{Our findings support \textbf{H2a}, indicating that these techniques were more effective than the Baseline in reducing both physical and time demands. Regarding \textbf{H2b}, however, only the Rear Passthrough AR (RPAR) technique demonstrated a higher level of immersion compared to the Baseline, while the other two techniques did not show a similar effect. As for \textbf{H2c}, our results provide partial support, since while none of the three techniques outperformed the Baseline in terms of overall motion sickness scores, they all helped mitigate disorientation. Previous research \cite{walker2010head, ragan2016amplified, palmisano2023differences} and our Study 1 have highlighted that turning the head is effortful, troublesome, reduces immersion, and can induce motion sickness. Our Study 2 has confirmed that, apart from not fully addressing the immersion issue, presenting the rear view in front of the user can, to some extent, improve other aspects.}

Among the three systems, Full Rear Passthrough+Pause (FRPP) and Rear Passthrough AR (RPAR) were the most preferred by participants, with FRPP showing a slight overall advantage. Subjective measurements revealed that FRPP outperformed the Baseline in reducing frustration, improving urgency handling, providing useful information, facilitating communication, and lowering effort and motion sickness scores. These benefits likely stem from its ability to display full rear awareness from the front and pause the gameplay, allowing users to focus better on observing the rear situation. RPAR was the second most preferred by participants; compared to the Baseline, it reduced frustration and provided the highest level of immersion and engagement, although it imposed a higher mental demand. These findings suggest that, when addressing rear awareness, users are generally more willing to switch to the full Passthrough view to handle interruptions, even if it means pausing the game. Previous studies \cite{o2023re, von2019you} found that pausing for a full Passthrough view rarely offered benefits over Passthrough AR, as pauses were seen as disruptive. Passthrough AR is effective for front awareness, but rear awareness requires more effort, making pauses less objectionable.

% Previous studies \cite{o2023re, von2019you} on displaying front views using Passthrough found that pausing the VR application to switch to a full Passthrough view rarely provided any benefit compared to using Passthrough AR, as the pauses were often perceived as disruptive. Many studies \cite{o2023re, gottsacker2021diegetic, von2019you} have noted that users generally prefer not to pause VR applications when dealing with real-world interruptions. Passthrough AR is effective at providing front awareness, allowing users to seamlessly see both real and virtual elements without affecting their gameplay. However, rear awareness is different from front awareness; observing the front feels more natural, while perceiving the rear through a front view is less common and requires extra mental effort. Consequently, pausing the game in this context is often less objectionable and can even be valued in certain situations.

Rear Passthrough Window (RPW) was less effective, as participants found the window format mentally demanding and not well integrated with the virtual environment. Guo et al. \cite{guo2024enhancement} also noted the criticism of windowed Passthrough, although their study did not find an increased mental load. Our study suggests that a window format is ineffective for both rear and front awareness.

% In contrast, Rear Passthrough Window (RPW), was not considered an effective system for displaying rear awareness. Participants found the window format mentally demanding and felt it did not blend well with the virtual environment. Similarly, Guo et al. \cite{guo2024enhancement} reported that participants criticized the unnatural integration of a windowed Passthrough view designed to help them see nearby bystanders. However, their study did not identify increased mental load from the windowed format, possibly because participants did not need to frequently monitor the window's content. In contrast, our study required users to focus on the real world within a short time frame. Overall, whether for rear or front awareness, presenting Passthrough views in a window format does not appear to be an effective approach.

\section{Summary of Findings and Design Insights}
Combining the findings from these two studies, we observed that when interruptions come from the rear, users' overall willingness to use Passthrough to view these interruptions is lower compared to when interruptions come from the front or sides. Our study addresses a gap in previous research by being the first to explore how Passthrough-based techniques can be used to display rear-facing interruptions in front of the user. The three techniques we tested improved user experience and reduced motion sickness to some extent compared to the Baseline head-turning condition. Among these methods, both FRPP and RPAR are considered effective, depending on whether users are willing to pause their game. In contrast, presenting the rear view in a window format (RPW) is not recommended, as it requires higher mental effort and is perceived as integrating unnaturally with the virtual environment.

% Presenting the rear view in front of the user proves to be a viable approach, offering notable advantages over the commonly used Baseline, which requires head-turning. This method can reduce physical and time demands, mitigate potential motion sickness risks, and enhance the overall experience. These benefits underscore the effectiveness of using these Passthrough techniques to handle interruptions from the rear.

\textcolor{black}{However, there are still challenges in applying these techniques in real-world settings. First, our experiment focused primarily on interruptions from the rear, without considering scenarios where interruptions may occur from a slightly more lateral direction. One potential solution could involve adding additional cameras or using a 360-degree camera to capture input from multiple directions \cite{anthes2016state} and then presenting the resulting Passthrough views from the front using the techniques we have proposed in this work. Moreover, based on findings from Study 1, participants indicated that slightly rotating their heads to observe interruptions from the side was acceptable. This approach could also be applied to rear interruptions. In practice, the system would need to integrate head tracking to align with the user's head movements and display the appropriate view accordingly \cite{wu2019effect, adhanom2023eye}. To align with users' common experience of using rear-view mirrors and reduce cognitive load, we adopted a mirrored design for the views. However, this could affect how the corresponding Passthrough view is presented during head movements, and further investigation is needed to determine the optimal display mode to prevent confusion or motion sickness.}

\textcolor{black}{Secondly, allowing users to choose whether to display the front or rear Passthrough view introduces the need for an additional input, which should be carefully designed to minimize user effort. Previous studies \cite{tidwell2010designing, stone2005user} suggest that when adding new features, it is important to consider existing similar features to align with user habits and reduce the learning curve, while also ensuring clear distinctions to avoid confusion. Furthermore, switching between the two views or removing the headset after viewing the rear Passthrough might lead to disorientation. In Study 2, although most participants did not report this issue, two participants mentioned that they felt confused. Since it is uncommon to view the real world from the rear in this way, additional cues or prompts might be required \cite{medeiros2021promoting} to reduce confusion and help users navigate these transitions smoothly.}

% Based on Study 2, we have developed design recommendations for presenting rear Passthrough views in front of the user. Both FRPP and RPAR are deemed effective methods, depending on whether users are willing to pause their game. These techniques demonstrate several advantages over the Baseline. However, since observing the rear view from the front is an unconventional experience, RPAR may increase mental load, making pausing the game beneficial for better observation. Conversely, presenting the rear view in a window format (RPW) is not recommended, as it demands higher mental effort and is considered to integrate unnaturally with the virtual environment. These results highlight the differences between presenting rear views and front views, offering valuable insights for future design considerations.

\section{Limitations and Future Work}
While this research offers valuable insights into enhancing users' rear awareness in VR, several limitations could be explored in future studies. Firstly, to ensure consistent participant experiences and better control of variables, we used pre-prepared materials to simulate interruption events. Future research could involve using cameras mounted behind the headset to capture real-world interruptions, allowing for the assessment of these techniques in practical applications.

As our study focused on presenting rear awareness to users, we only evaluated scenarios where interruptions originated from the rear. In real-life usage, users may need to respond to interruptions from different directions, requiring them to switch between front and rear Passthrough views. Future studies could explore how such situations, where the two views are employed, affect user experience.

Viewing the rear view from the front is unconventional, and participants in our experiment were exposed to it only briefly. Some noted that with prolonged use, they would likely become more accustomed to this viewing method. Therefore, future research could explore this further through long-term studies to better understand and address the potential mental load associated with this approach. \textcolor{black}{Additionally, the game used in our experiment, Fruit Ninja, is a fast-paced action game. The results may differ when players or users engage with games or other types of applications that do not emphasize time, such as casual games, and this also warrants further investigation in future research.}

Furthermore, in Study 2, the interruptions were all perceived by participants through auditory cues, reflecting the majority of real-world interruptions users experience in daily life. However, interruptions can also be triggered by other sensory cues, such as tactile sensations—like feeling a collision—or olfactory cues, such as detecting a smell. Future research could explore the effectiveness of using Passthrough techniques to handle interruptions triggered by these types of cues.

\section{Conclusion}
This work indicates that the direction of real-world interruptions significantly influences the use of Passthrough, especially when interruptions come from the rear, as this requires additional head-turning, which is often seen as troublesome, effortful, and potentially induces motion sickness, leading to a negative experience. To address this, we proposed three techniques that present the rear Passthrough view in front of the user. Our findings confirm the effectiveness of these techniques, demonstrating clear advantages over the conventional method of turning the head and using Passthrough, including reduced physical demands, time, motion sickness, and an overall improved user experience. Among the systems evaluated, Full Rear Passthrough + Pause (FRPP) and Rear Passthrough
AR (RPAR) are the most preferred, with FRPP showing a slight overall edge. Unlike the findings from previous studies on front awareness, we found that users were not opposed to pausing gameplay when perceiving rear awareness, likely because the unconventional perspective required additional mental effort. These results highlight the benefits of providing rear awareness in the front and offer valuable insights for future VR design.

\bibliographystyle{abbrv-doi}

\bibliography{template}

\begin{thebibliography}{10}

\bibitem{adhanom2023eye}
I.~B. Adhanom, P.~MacNeilage, and E.~Folmer.
\newblock Eye tracking in virtual reality: a broad review of applications and challenges.
\newblock {\em Virtual Reality}, 27(2):1481--1505, 2023.

\bibitem{anthes2016state}
C.~Anthes, R.~J. Garc{\'\i}a-Hern{\'a}ndez, M.~Wiedemann, and D.~Kranzlm{\"u}ller.
\newblock State of the art of virtual reality technology.
\newblock In {\em 2016 IEEE aerospace conference}, pp. 1--19. IEEE, 2016.

\bibitem{bailenson2024seeing}
J.~N. Bailenson, B.~Beams, J.~Brown, C.~DeVeaux, E.~Han, A.~C. Queiroz, R.~Ratan, M.~Santoso, T.~Srirangarajan, Y.~Tao, et~al.
\newblock Seeing the world through digital prisms: Psychological implications of passthrough video usage in mixed reality.
\newblock 2024.

\bibitem{banquiero2023passthrough}
M.~Banquiero, G.~Valdeolivas, S.~Trincado, N.~Garc{\'\i}a, and M.~C. Juan.
\newblock Passthrough mixed reality with oculus quest 2: A case study on learning piano.
\newblock {\em IEEE MultiMedia}, 2023.

\bibitem{barai2020outside}
S.~Barai and M.~Momin.
\newblock Outside-in electromagnetic tracking method for augmented and virtual reality 6-degree of freedom head-mounted displays.
\newblock In {\em 2020 4th International Conference on Intelligent Computing and Control Systems (ICICCS)}, pp. 467--476. IEEE, 2020.

\bibitem{bowman2023using}
R.~Bowman, C.~Nadal, K.~Morrissey, A.~Thieme, and G.~Doherty.
\newblock Using thematic analysis in healthcare hci at chi: A scoping review.
\newblock In {\em Proceedings of the 2023 CHI Conference on Human Factors in Computing Systems}, pp. 1--18, 2023.

\bibitem{caserman2021cybersickness}
P.~Caserman, A.~Garcia-Agundez, A.~G{\'a}mez~Zerban, and S.~G{\"o}bel.
\newblock Cybersickness in current-generation virtual reality head-mounted displays: systematic review and outlook.
\newblock {\em Virtual Reality}, 25(4):1153--1170, 2021.

\bibitem{cobb1999measurement}
S.~V.~G. Cobb.
\newblock Measurement of postural stability before and after immersion in a virtual environment.
\newblock {\em Applied ergonomics}, 30(1):47--57, 1999.

\bibitem{coutrix2024impact}
C.~Coutrix and C.~Prost.
\newblock Impact of fingernails length on mobile tactile interaction.
\newblock In {\em Proceedings of the CHI Conference on Human Factors in Computing Systems}, pp. 1--21, 2024.

\bibitem{dao2021bad}
E.~Dao, A.~Muresan, K.~Hornb{\ae}k, and J.~Knibbe.
\newblock Bad breakdowns, useful seams, and face slapping: Analysis of vr fails on youtube.
\newblock In {\em Proceedings of the 2021 chi conference on human factors in computing systems}, pp. 1--14, 2021.

\bibitem{feld2024simple}
N.~Feld, P.~Bimberg, B.~Weyers, and D.~Zielasko.
\newblock Simple and efficient? evaluation of transitions for task-driven cross-reality experiences.
\newblock {\em IEEE Transactions on Visualization and Computer Graphics}, 2024.

\bibitem{freiwald2018camera}
J.~P. Freiwald, N.~Katzakis, and F.~Steinicke.
\newblock Camera time warp: compensating latency in video see-through head-mounted-displays for reduced cybersickness effects.
\newblock In {\em Proceedings of the 24th ACM symposium on virtual reality software and technology}, pp. 1--7, 2018.

\bibitem{george2018intelligent}
C.~George, M.~Demmler, and H.~Hussmann.
\newblock Intelligent interruptions for ivr: investigating the interplay between presence, workload and attention.
\newblock In {\em Extended abstracts of the 2018 CHI conference on human factors in computing systems}, pp. 1--6, 2018.

\bibitem{george2019should}
C.~George, P.~Janssen, D.~Heuss, and F.~Alt.
\newblock Should i interrupt or not? understanding interruptions in head-mounted display settings.
\newblock In {\em Proceedings of the 2019 on designing interactive systems conference}, pp. 497--510, 2019.

\bibitem{george2020seamless}
C.~George, A.~N. Tien, and H.~Hussmann.
\newblock Seamless, bi-directional transitions along the reality-virtuality continuum: A conceptualization and prototype exploration.
\newblock In {\em 2020 IEEE International Symposium on Mixed and Augmented Reality (ISMAR)}, pp. 412--424. IEEE, 2020.

\bibitem{ghosh2018notifivr}
S.~Ghosh, L.~Winston, N.~Panchal, P.~Kimura-Thollander, J.~Hotnog, D.~Cheong, G.~Reyes, and G.~D. Abowd.
\newblock Notifivr: Exploring interruptions and notifications in virtual reality.
\newblock {\em IEEE transactions on visualization and computer graphics}, 24(4):1447--1456, 2018.

\bibitem{gottsacker2021diegetic}
M.~Gottsacker, N.~Norouzi, K.~Kim, G.~Bruder, and G.~Welch.
\newblock Diegetic representations for seamless cross-reality interruptions.
\newblock In {\em 2021 IEEE International Symposium on Mixed and Augmented Reality (ISMAR)}, pp. 310--319. IEEE, 2021.

\bibitem{gottsacker2022exploring}
M.~Gottsacker, R.~Syamil, P.~Wisniewski, G.~Bruder, C.~Cruz-Neira, and G.~Welch.
\newblock Exploring cues and signaling to improve cross-reality interruptions.
\newblock In {\em 2022 IEEE International Symposium on Mixed and Augmented Reality Adjunct (ISMAR-Adjunct)}, pp. 827--832. IEEE, 2022.

\bibitem{graybiel1968rapid}
A.~Graybiel and C.~D. Wood.
\newblock {\em Rapid vestibular adaptation in a rotating environment by means of controlled head movements}, vol. 1053.
\newblock Naval Aerospace Medical Institute, Naval Aerospace Medical Center, 1968.

\bibitem{guest2006many}
G.~Guest, A.~Bunce, and L.~Johnson.
\newblock How many interviews are enough? an experiment with data saturation and variability.
\newblock {\em Field methods}, 18(1):59--82, 2006.

\bibitem{guo2024exploring}
Z.~Guo, H.~D. Deng, H.~Wang, A.~Tan, W.~Xu, and H.-N. Liang.
\newblock Exploring the impact of passthrough on vr exergaming in public environments: A field study.
\newblock In {\em 2024 IEEE International Symposium on Mixed and Augmented Reality (ISMAR)}. IEEE, 2024.

\bibitem{guo2024breaking}
Z.~Guo, H.~Wang, H.~Deng, W.~Xu, N.~Baghaei, C.-H. Lo, and H.-N. Liang.
\newblock Breaking the isolation: Exploring the impact of passthrough in shared spaces on player performance and experience in vr exergames.
\newblock {\em IEEE Transactions on Visualization and Computer Graphics}, 2024.

\bibitem{guo2024enhancement}
Z.~Guo, W.~Xu, H.~Wang, T.~Wan, N.~Baghaei, C.-H. Lo, and H.-N. Liang.
\newblock Enhancement of co-located shared vr experiences: Representing non-hmd observers on both hmd and 2d screens.
\newblock In {\em 2024 IEEE International Symposium on Mixed and Augmented Reality (ISMAR)}. IEEE, 2024.

\bibitem{hanington2019universal}
B.~Hanington and B.~Martin.
\newblock {\em Universal methods of design expanded and revised: 125 Ways to research complex problems, develop innovative ideas, and design effective solutions}.
\newblock Rockport publishers, 2019.

\bibitem{hart2006nasa}
S.~G. Hart.
\newblock Nasa-task load index (nasa-tlx); 20 years later.
\newblock In {\em Proceedings of the human factors and ergonomics society annual meeting}, vol.~50, pp. 904--908. Sage publications Sage CA: Los Angeles, CA, 2006.

\bibitem{hartmann2019realitycheck}
J.~Hartmann, C.~Holz, E.~Ofek, and A.~D. Wilson.
\newblock Realitycheck: Blending virtual environments with situated physical reality.
\newblock In {\em Proceedings of the 2019 CHI Conference on Human Factors in Computing Systems}, pp. 1--12, 2019.

\bibitem{kanamori2018obstacle}
K.~Kanamori, N.~Sakata, T.~Tominaga, Y.~Hijikata, K.~Harada, and K.~Kiyokawa.
\newblock Obstacle avoidance method in real space for virtual reality immersion.
\newblock In {\em 2018 IEEE International Symposium on Mixed and Augmented Reality (ISMAR)}, pp. 80--89. IEEE, 2018.

\bibitem{kennedy1993simulator}
R.~S. Kennedy, N.~E. Lane, K.~S. Berbaum, and M.~G. Lilienthal.
\newblock Simulator sickness questionnaire: An enhanced method for quantifying simulator sickness.
\newblock {\em The international journal of aviation psychology}, 3(3):203--220, 1993.

\bibitem{kern2003context}
N.~Kern and B.~Schiele.
\newblock Context-aware notification for wearable computing.
\newblock In {\em Seventh IEEE International Symposium on Wearable Computers, 2003. Proceedings.}, pp. 223--223. IEEE Computer Society, 2003.

\bibitem{keshavarz2011axis}
B.~Keshavarz and H.~Hecht.
\newblock Axis rotation and visually induced motion sickness: the role of combined roll, pitch, and yaw motion.
\newblock {\em Aviation, space, and environmental medicine}, 82(11):1023--1029, 2011.

\bibitem{kim2022mobile}
J.~Kim, Y.~Choi, M.~Xia, and J.~Kim.
\newblock Mobile-friendly content design for moocs: challenges, requirements, and design opportunities.
\newblock In {\em Proceedings of the 2022 CHI Conference on Human Factors in Computing Systems}, pp. 1--16, 2022.

\bibitem{kudo2021towards}
Y.~Kudo, A.~Tang, K.~Fujita, I.~Endo, K.~Takashima, and Y.~Kitamura.
\newblock Towards balancing vr immersion and bystander awareness.
\newblock {\em Proceedings of the ACM on Human-Computer Interaction}, 5(ISS):1--22, 2021.

\bibitem{kudry2023enhanced}
P.~Kudry and M.~Cohen.
\newblock Enhanced wearable force-feedback mechanism for free-range haptic experience extended by pass-through mixed reality.
\newblock {\em Electronics}, 12(17):3659, 2023.

\bibitem{kuo2023reprojection}
G.~Kuo, E.~Penner, S.~Moczydlowski, A.~Ching, D.~Lanman, and N.~Matsuda.
\newblock Reprojection-free vr passthrough.
\newblock In {\em ACM SIGGRAPH 2023 Emerging Technologies}, SIGGRAPH '23. Association for Computing Machinery, New York, NY, USA, 2023. doi: {{%
10\hspace{.1pt}\discretionary{.}{%
}{.}\hspace{.4pt}1145\discretionary{/}{%
}{/}3588037\hspace{.1pt}\discretionary{.}{%
}{.}\hspace{.4pt}3595391}}


\bibitem{langbehn2019turn}
E.~Langbehn, J.~Wittig, N.~Katzakis, and F.~Steinicke.
\newblock Turn your head half round: Vr rotation techniques for situations with physically limited turning angle.
\newblock In {\em Proceedings of Mensch und Computer 2019}, pp. 235--243. 2019.

\bibitem{lawson2016neurovestibular}
B.~D. Lawson, A.~H. Rupert, and B.~J. McGrath.
\newblock The neurovestibular challenges of astronauts and balance patients: some past countermeasures and two alternative approaches to elicitation, assessment and mitigation.
\newblock {\em Frontiers in systems neuroscience}, 10:96, 2016.

\bibitem{livatino2022effects}
S.~Livatino, A.~Zocco, Y.~Iqbal, P.~Gainley, G.~Morana, and G.~M. Farinella.
\newblock Effects of head rotation and depth enhancement in virtual reality user-scene interaction.
\newblock In {\em International Conference on Extended Reality}, pp. 139--146. Springer, 2022.

\bibitem{mcgill2015dose}
M.~McGill, D.~Boland, R.~Murray-Smith, and S.~Brewster.
\newblock A dose of reality: Overcoming usability challenges in vr head-mounted displays.
\newblock In {\em Proceedings of the 33rd annual ACM conference on human factors in computing systems}, pp. 2143--2152, 2015.

\bibitem{mcgill2017passenger}
M.~McGill, A.~Ng, and S.~Brewster.
\newblock I am the passenger: how visual motion cues can influence sickness for in-car vr.
\newblock In {\em Proceedings of the 2017 chi conference on human factors in computing systems}, pp. 5655--5668, 2017.

\bibitem{medeiros2021promoting}
D.~Medeiros, R.~Dos~Anjos, N.~Pantidi, K.~Huang, M.~Sousa, C.~Anslow, and J.~Jorge.
\newblock Promoting reality awareness in virtual reality through proxemics.
\newblock In {\em 2021 IEEE Virtual Reality and 3D User Interfaces (VR)}, pp. 21--30. IEEE, 2021.

\bibitem{nwobodo2023review}
O.~J. Nwobodo, K.~Wereszczy{\'n}ski, and K.~Cyran.
\newblock A review on tracking head movement in augmented reality systems.
\newblock {\em Procedia Computer Science}, 225:4344--4353, 2023.

\bibitem{o2022exploring2}
J.~O'Hagan, M.~Khamis, M.~McGill, and J.~R. Williamson.
\newblock Exploring attitudes towards increasing user awareness of reality from within virtual reality.
\newblock In {\em Proceedings of the 2022 ACM International Conference on Interactive Media Experiences}, pp. 151--160, 2022.

\bibitem{o2020reality}
J.~O'Hagan and J.~R. Williamson.
\newblock Reality aware vr headsets.
\newblock In {\em Proceedings of the 9th ACM international symposium on pervasive displays}, pp. 9--17, 2020.

\bibitem{o2022exploring}
J.~O'Hagan, J.~R. Williamson, M.~Khamis, and M.~McGill.
\newblock Exploring manipulating in-vr audio to facilitate verbal interactions between vr users and bystanders.
\newblock In {\em Proceedings of the 2022 International Conference on Advanced Visual Interfaces}, pp. 1--9, 2022.

\bibitem{o2023re}
J.~O'Hagan, J.~R. Williamson, F.~Mathis, M.~Khamis, and M.~McGill.
\newblock Re-evaluating vr user awareness needs during bystander interactions.
\newblock In {\em Proceedings of the 2023 CHI Conference on Human Factors in Computing Systems}, pp. 1--17, 2023.

\bibitem{o2021safety}
J.~O’Hagan, J.~R. Williamson, M.~McGill, and M.~Khamis.
\newblock Safety, power imbalances, ethics and proxy sex: Surveying in-the-wild interactions between vr users and bystanders.
\newblock In {\em 2021 IEEE International Symposium on Mixed and Augmented Reality (ISMAR)}, pp. 211--220. IEEE, 2021.

\bibitem{palmisano2023differences}
S.~Palmisano, R.~S. Allison, J.~Teixeira, and J.~Kim.
\newblock Differences in virtual and physical head orientation predict sickness during active head-mounted display-based virtual reality.
\newblock {\em Virtual Reality}, 27(2):1293--1313, 2023.

\bibitem{pointecker2022bridging}
F.~Pointecker, J.~Friedl, D.~Schwajda, H.-C. Jetter, and C.~Anthes.
\newblock Bridging the gap across realities: Visual transitions between virtual and augmented reality.
\newblock In {\em 2022 IEEE international symposium on mixed and augmented reality (ISMAR)}, pp. 827--836. IEEE, 2022.

\bibitem{ragan2016amplified}
E.~D. Ragan, S.~Scerbo, F.~Bacim, and D.~A. Bowman.
\newblock Amplified head rotation in virtual reality and the effects on 3d search, training transfer, and spatial orientation.
\newblock {\em IEEE transactions on visualization and computer graphics}, 23(8):1880--1895, 2016.

\bibitem{rzayev2019notification}
R.~Rzayev, S.~Mayer, C.~Krauter, and N.~Henze.
\newblock Notification in vr: The effect of notification placement, task and environment.
\newblock In {\em Proceedings of the annual symposium on computer-human interaction in play}, pp. 199--211, 2019.

\bibitem{santoso2024video}
M.~Santoso and J.~N. Bailenson.
\newblock How video passthrough headsets influence perception of self and others.
\newblock {\em arXiv preprint arXiv:2407.16904}, 2024.

\bibitem{sargunam2017guided}
S.~P. Sargunam, K.~R. Moghadam, M.~Suhail, and E.~D. Ragan.
\newblock Guided head rotation and amplified head rotation: Evaluating semi-natural travel and viewing techniques in virtual reality.
\newblock In {\em 2017 IEEE Virtual Reality (VR)}, pp. 19--28. IEEE, 2017.

\bibitem{scavarelli2017vr}
A.~Scavarelli and R.~J. Teather.
\newblock Vr collide! comparing collision-avoidance methods between co-located virtual reality users.
\newblock In {\em Proceedings of the 2017 CHI conference extended abstracts on human factors in computing systems}, pp. 2915--2921, 2017.

\bibitem{schubert2001experience}
T.~Schubert, F.~Friedmann, and H.~Regenbrecht.
\newblock The experience of presence: Factor analytic insights.
\newblock {\em Presence: Teleoperators \& Virtual Environments}, 10(3):266--281, 2001.

\bibitem{stone2005user}
D.~Stone, C.~Jarrett, M.~Woodroffe, and S.~Minocha.
\newblock {\em User interface design and evaluation}.
\newblock Elsevier, 2005.

\bibitem{tidwell2010designing}
J.~Tidwell.
\newblock {\em Designing interfaces: Patterns for effective interaction design}.
\newblock " O'Reilly Media, Inc.", 2010.

\bibitem{ujike2004effects}
H.~Ujike, T.~Yokoi, and S.~Saida.
\newblock Effects of virtual body motion on visually-induced motion sickness.
\newblock In {\em The 26th Annual International Conference of the IEEE Engineering in Medicine and Biology Society}, vol.~1, pp. 2399--2402. IEEE, 2004.

\bibitem{von2019you}
J.~Von~Willich, M.~Funk, F.~M{\"u}ller, K.~Marky, J.~Riemann, and M.~M{\"u}hlh{\"a}user.
\newblock You invaded my tracking space! using augmented virtuality for spotting passersby in room-scale virtual reality.
\newblock In {\em Proceedings of the 2019 on Designing Interactive Systems Conference}, pp. 487--496, 2019.

\bibitem{walker2010head}
A.~D. Walker, E.~R. Muth, F.~S. Switzer, and A.~Hoover.
\newblock Head movements and simulator sickness generated by a virtual environment.
\newblock {\em Aviation, space, and environmental medicine}, 81(10):929--934, 2010.

\bibitem{wang2022realitylens}
C.-H. Wang, B.-Y. Chen, and L.~Chan.
\newblock Realitylens: A user interface for blending customized physical world view into virtual reality.
\newblock In {\em Proceedings of the 35th Annual ACM Symposium on User Interface Software and Technology}, pp. 1--11, 2022.

\bibitem{wang2022real}
J.~Wang, H.-N. Liang, D.~Monteiro, W.~Xu, and J.~Xiao.
\newblock Real-time prediction of simulator sickness in virtual reality games.
\newblock {\em IEEE Transactions on Games}, 15(2):252--261, 2022.

\bibitem{wang2024omnidirectional}
J.~Wang, R.~Shi, X.~Li, Y.~Wei, and H.-N. Liang.
\newblock Omnidirectional virtual visual acuity: A user-centric visual clarity metric for virtual reality head-mounted displays and environments.
\newblock {\em IEEE Transactions on Visualization and Computer Graphics}, 2024.

\bibitem{wang2023effect}
J.~Wang, R.~Shi, W.~Zheng, W.~Xie, D.~Kao, and H.-N. Liang.
\newblock Effect of frame rate on user experience, performance, and simulator sickness in virtual reality.
\newblock {\em IEEE Transactions on Visualization and Computer Graphics}, 29(5):2478--2488, 2023.

\bibitem{warburton2023measuring}
M.~Warburton, M.~Mon-Williams, F.~Mushtaq, and J.~R. Morehead.
\newblock Measuring motion-to-photon latency for sensorimotor experiments with virtual reality systems.
\newblock {\em Behavior research methods}, 55(7):3658--3678, 2023.

\bibitem{wobbrock2011aligned}
J.~O. Wobbrock, L.~Findlater, D.~Gergle, and J.~J. Higgins.
\newblock The aligned rank transform for nonparametric factorial analyses using only anova procedures.
\newblock In {\em Proceedings of the SIGCHI conference on human factors in computing systems}, pp. 143--146, 2011.

\bibitem{wu2019effect}
T.~L. Wu, A.~Gomes, K.~Fernandes, and D.~Wang.
\newblock The effect of head tracking on the degree of presence in virtual reality.
\newblock {\em International Journal of Human--Computer Interaction}, 35(17):1569--1577, 2019.

\bibitem{xiong2024petpresence}
N.~Xiong, Q.~Liu, and K.~Zhu.
\newblock Petpresence: Investigating the integration of real-world pet activities in virtual reality.
\newblock {\em IEEE Transactions on Visualization and Computer Graphics}, 2024.

\end{thebibliography}
\end{document}